\documentstyle[12pt]{article}
\newcommand{\B}{B1422$+$231}

\newcommand{\kms}{km\,s$^{-1}$}
\newcommand{\kmsMpc}{km\,s$^{-1}$\,Mpc$^{-1}$}

\newcommand{\cf}{cf.}
\newcommand{\eg}{e.g.}
\newcommand{\etal}{et~al.}

\newcommand{\ie}{i.e.}

\newcommand{\balpha}{\mbox{\boldmath $\alpha$}}	   
\newcommand{\bbeta}{\mbox{\boldmath $\beta$}}	   
\newcommand{\btheta}{\mbox{\boldmath $\theta$}}	   

\title{The Gravitational Lens System B1422$+$231:
Dark Matter, Superluminal Expansion and the Hubble Constant}
\author{David W. Hogg \& R. D. Blandford \\
{\em 130-33 Caltech, Pasadena, CA 91125.} }
\date{submitted 1993 July 23, revised 1993 December 1}

\begin{document}


\maketitle

\begin{abstract}
A gravitational lens model of the radio quasar \B\ is presented which
can account for the image arrangement and approximately for the
relative magnifications.  The locations of the principal lensing mass
and a more distant secondary mass concentration were predicted and
subsequently luminous galaxies were found at these locations.  This
argues against the existence of substantial numbers of ``dark''
galaxies.  The model suggests that if the compact radio source is
intrinsically superluminal then the observed component motions may be
as large as $\sim100c$ with image B moving in the opposite direction
to images A and C.  The prospects for a measuring the Hubble constant
from a model incorporating lens galaxy locations, compact radio source
expansion speeds and radio time delays, if and when these are
measured, are briefly assessed.

\vspace{2ex}

\B---quasars: general---dark matter---distance scale.
\end{abstract}

\section{Introduction}

Gravitational lensing provides a direct probe of the mass
distributions in astronomical objects.  It is therefore important to
find accurate models of lens potentials.  Indeed, the existence of a
compelling lens model can provide corroboration of multiple imaging.
In addition, an accurate, verifiable model opens the possibility of
measuring the Hubble constant if the source proves to be variable.

The \B\ system was discovered by Patnaik \etal\ (1992) as part of a
survey of flat-spectrum radio sources.  It has a redshift of $z=3.62$,
and consists of three bright components (A, B and C) and one dim
component (D) all within 1.3 arcseconds.  There is no evidence, as
yet, of extended emission.  Although D is too dim for accurate
spectral and polarization measurements, A, B and C have similar radio
spectra and fractional polarization.  The location of D is consistent
with the hypothesis that it is the fourth image of a gravitational
lens system.  Lawrence \etal\ (1992) observed \B\ in the infrared and
found that all four components are in similar positions and have
roughly the same flux ratios as in the radio.  For these reasons, both
sets of authors conclude that \B\ is a gravitational lens system.

Simple gravitational-lens models of \B\ predicted a mass concentration
inside the circle of images, plus a mass concentration to the
southeast (Hogg \& Blandford 1993).  Recently several galaxies were
discovered in or near the \B\ system.  Yee \& Ellingson (1993) have
found a galaxy (G1) inside the circle of images, and Larkin \etal\
(1993) have found two more galaxies (G2, G3) to the southeast of the
quasar.

In this paper, we describe a simple lens model which provides an
adequate description of existing observational data and predicts the
velocity dispersions or masses of the lensing objects, the relative
directions and approximate relative expansion speeds of the compact
radio components (if they turn out to be intrinsically superluminal),
and the relative time delays between intensity variations in the
images.  We find that using the nearby galaxy locations in our model
significantly improves our fits, suggesting that the recently
discovered galaxies are the main contributors to the lensing
potential.  Future observations (\eg\ with {\sc vlbi}) should improve
our ability to make accurate models, and thereby improve the remaining
predictions.

\section{Observations}

The source \B\ was observed by Patnaik \etal\ (1992) with the {\sc
vla} at 8.4~GHz, and with the {\sc merlin} array at 5~GHz (Table 1).
The total flux density is roughly 0.5~Jy at 5~GHz, with $\sim 3\%$
polarization.  In the optical, the combined source has the spectrum of
a $16.5^{\rm m}$ luminous, red quasar, with absolute magnitude $M_V=
-29.5+5\log h$ (where $h$ is the Hubble constant in units of
100~\kmsMpc, and $\Omega_0=1$).  Observations by Lawrence \etal\
(1992) in the 2.0--$2.4\mu$ $K$ band made with the Caltech
$58\times62$ InSb array camera at the Cassegrain focus of the Hale
Telescope are consistent with the radio results.

The primary galaxy G1 was observed by Yee \& Ellingson (1993) with the
{\sc cfht}.  When the four known components of the \B\ system were
subtracted, the galaxy G1 was found inside the circle of images.  G2
and G3 were observed by Larkin \etal\ (1993) on a $2\mu$ image taken
by the Keck Telescope.  These two galaxies are very bright; G2 and G3
have $K$-band magnitudes of 16.2 and 15.7 respectively.  The positions
of G1, G2, and G3 are given in Table~2.

A preliminary measurement of the redshift of G1 of $z=0.64$ was found
by Hammer \etal\ (1993).

\section{Lens model}

\subsection{Gravitational lens theory}

The {\em lens equation\/} relates image positions to source positions.
The potential of a gravitational lens can be projected onto a
two-dimensional plane (the {\em lens plane\/}) orthogonal to the
direction of light propagation and scaled so that the lens equation is
\begin{equation}
\bbeta(\btheta)=\btheta-\balpha(\btheta)=
\btheta-\nabla_{\!\btheta}\psi (\btheta)
\end{equation}
where $\bbeta$ (a two-dimensional vector) is the angular position of
the source (on the {\em source plane\/}), $\btheta$ is that of an
image point (on the {\em image plane\/}), $\balpha$ is the reduced
deflection, $\nabla_{\!\btheta}$ is the two-dimensional
gradient operator with respect to $\btheta$, and $\psi(\btheta)$ is
the {\em deflection potential\/} (\eg\ Schneider, Ehlers \& Falco
1992; Blandford \& Narayan 1992).

The inverse of the Jacobian matrix of the mapping $\bbeta(\btheta)$
from the image plane to the source plane is the {\em magnification
tensor\/} $[\mu(\btheta)]$,
\begin{equation}
\left[\mu(\btheta)\right]
=\left[\frac{\partial\bbeta}{\partial\btheta}\right]^{-1} .
\end{equation}
The (scalar) magnification $\mu$ of an image relative to the source is
the determinant of the magnification tensor.

The deflection potential $\psi$ is related to the ``surface'' mass
density (mass per unit solid angle) $\Sigma$ by
\begin{equation}
\nabla^2\psi=\frac{2\Sigma}{\Sigma_c}, \;\;\; {\rm where} \;\;\;
\Sigma_c=\frac{c^2}{4\pi G}\frac{D_{\rm d}D_{\rm s}}{D_{\rm ds}},
\end{equation}
is the critical density and
where $D_{\rm d}$, $D_{\rm s}$, and $D_{\rm ds}$ are the angular
diameter distances from observer to lens (``deflector''), observer to
source, and lens to source.

There is a time delay $\Delta t$ associated with an image at $\btheta$
from a source at $\bbeta$ given by
\begin{equation}
\Delta t=
T_0\left[\frac{1}{2}\left|\btheta-\bbeta\right|^2-\psi (\btheta)\right] ,
\;\;\; {\rm where} \;\;\;
T_0=(1+z_{\rm d})\frac{D_{\rm d}D_{\rm s}}{cD_{\rm ds}}
\end{equation}
where $z_{\rm d}$ is the redshift of the lensing potential.  By
Fermat's principle the images are located at stationary values of
$\Delta t(\btheta)$ (\eg\ Schneider \etal\ 1992).

\subsection{Fitting the data}

We use a general procedure to fit a gravitational lens model to the
data (\cf\ Kayser 1990; Kochanek 1991a).  In the case of \B, our data
comprise $n_O=12$ observations $\{O_i\}$, namely the positions and
fluxes of the four images.  We suppose that these observations are
normally distributed with variances $\sigma^2_i$ which we estimate
from the results of Patnaik \etal\ (1992).  We then consider lens
models characterized by $n_M$ model parameters denoted $\{M_j\}$.  The
number of degrees of freedom is $\nu=n_O-n_M$ (\cf\ Kochanek 1991b).
The model parameters include lens parameters plus three more
parameters which specify the location and intensity of the source on
the source plane.

We seek a best-fit solution by minimizing a least squares figure of
merit
\begin{equation}
\chi^2=\sum_{i=1}^{n_O}\frac{(O'_i(\{M_j\})-O_i)^2}{\sigma^2_i}
\end{equation}
with respect to variation of $\{M_j\}$, where the $O'_i(\{M_j\})$ are
the values for the $O_i$ that we derive from the model (\eg\ Press
\etal\ 1992).

Roughly speaking, a fit is good if $\chi^2/\nu$ is on the order of
unity.  Adding additional parameters to a model will almost always
allow a reduction in $\chi^2$, but it is only significant if it
reduces $\chi^2/\nu$.  This rule of thumb allows us to test the
significance of adding additional parameters.

Prior to performing the fit, it was necessary to explore model space
using the time delay surface.  We identified all the extrema and then
minimized a modified $\chi^2$ obtained by deriving a source location
and magnification for each image (by backwards ray-tracing), given a
model, and then using the magnification tensors to convert the source
displacements to image displacements for a given source location.
This is a fast procedure for searching large volumes of parameter
space.

Given a particular model which minimizes $\chi^2$ we form the Hessian
matrix,
\begin{equation}
D_{jj'}=\sum_{i=1}^{n_O}\frac{1}{\sigma_i^2}
\left(\frac{\partial O'_i}{\partial M_j}\right)
\left(\frac{\partial O'_i}{\partial M_{j'}}\right),
\end{equation}
numerically, by varying the model parameters.  The inverse of matrix
$D_{jj'}$ is the covariance matrix $C_{jj'}$ which provides standard
deviations for the parameters.

The goal of this model-fitting is to predict future observables, which
we denote by $\{P_k\}$.  In the present case these include the
relative magnification tensors of the three bright images and the time
delays for variations of all four images.  We compute the variances in
the future observables in the usual manner:
\begin{equation}
\sigma_k^2=\sum_{j,j'}\frac{\partial P_k}{\partial M_j}C_{jj'}
\frac{\partial P_k}{\partial M_{j'}}.
\end{equation}

\subsection{Lens potential models}

The history of this project is relevant to understanding the results.
We first found that a lens model based upon a single elliptical galaxy
is insufficient to account for the large magnification ratio between
images A, B, C and image D.  The simplest successful lens model, which
we call the {\em initial model,\/} comprises two singular isothermal
spheres with a common redshift, each with potential
$\psi=b|\btheta-{\btheta}_c|$, where the parameter
$b=4\pi\sigma^2D_{\rm ds}/c^2D_{\rm s}$, the critical radius, measures
the 1D velocity dispersion $\sigma$, and ${\btheta}_c$ locates the
centre of the potential (\eg\ Kochanek 1991a; Schneider \etal\ 1992).
(In practice, having a second potential breaks the circular symmetry
and avoids the structural instability of its imaging properties.) The
initial model has $n_M=9$ parameters, so $\nu=3$.

On the basis of this model (and simple variants) we predicted that
there should be a primary lensing galaxy located about $1/3$ of the
way from image D to image B, and that there should be a larger mass
located about $7''$ southeast of image B with equivalent velocity
dispersion $\sim 460$~\kms; presumably a small group (Hogg \&
Blandford 1993).  We then learned that Yee \& Ellingson (1993) found
the primary galaxy at the predicted location, Hammer \etal\ (1993)
made a preliminary measurement of the G1 redshift, and Larkin \etal\
(1993) discovered the two galaxies G2 and G3 to the southeast near the
predicted location.

We therefore explored refined models in which we fixed the locations
and redshifts of the three galaxies G1, G2, and G3 (we assume that G2
and G3 are at the same redshift as G1), and adjusted their mass
distributions.  We refer to the best model found this way as the {\em
refined model.\/} In this case, the two distant galaxies G2 and G3 are
modelled as point masses, each with potential $\psi=b^2\ln
|\btheta-{\btheta}_c|$, where again ${\btheta}_c$ locates the centre
of the potential, and $b^2=M_g/\pi\Sigma_c$ measures the mass $M_g$ of
the galaxy (\eg\ Kochanek 1991a; Schneider \etal\ 1992).  Treating a
distant galaxy as a point mass is valid if its mass distribution is
roughly circular (on the sky) and none of it overlaps the circle of
images.  The refined model we settle on has $n_M=6$, $\nu=6$.

\section{Results}

\subsection{Lens model}

The initial model is the simplest capable of approximately reproducing
the observations.  The centre of the primary potential is located
inside the circle of images, and the secondary potential is south-east
of D.  The best-fit parameters together with the errors derived on the
basis of the model are given in Table~3.  In Table~4 we give the
computed magnification tensors and time delays for images A, C and D
relative to image B together.  The largest deficiency of this model is
its failure to reproduce adequately the relative magnification of
images A and B.

We experimented with several sets of adjustable parameters for the
refined model.  The simplest set consists of only three parameters: a
velocity dispersion $\sigma$ for G1, and masses $M_2$ and $M_3$ for G2
and G3.  In this case G1 is modelled as a singular isothermal sphere,
and G2 and G3 are treated as point masses.  This set of parameters
can be expanded by adding a shear to the G1 potential, changing the
radial dependence of the G1 potential, or by fine-tuning the centre of
the G1 potential (always remaining consistent with the observations of
Yee \& Ellingson (1993).

We found that introducing the simple, three-parameter refined model
reduced $\chi^2/\nu$ from $\sim 60$ (for the initial model) to $\sim
16$.  We found that adding parameters to the refined model, such as an
extra shear, a modified G1 radial dependence, or a fine-tuned G1
location, did not reduce $\chi^2/\nu$ any further.  We therefore
settled on the very simple three-parameter refined model.  The
best-fit parameters for the refined model are given in Table~3,
computed magnification tensors and time delays are given in Table~4.
Again, the largest deficiency of the model is in reproducing the A/B
magnification ratio.

The refined model has a $\chi^2/\nu$ which is still quite large ($\sim
16$), but it must be remembered that the $\chi^2$ value depends on the
uncertainties in the fluxes, which we have taken from the radio
observations alone.  The descrepancies between the radio and infrared
observations suggest that the uncertainties in the fluxes may be
higher than we have assumed.  Either a $\sim 20\%$ reduction in the
radio flux of A, or a factor of 2--3 increase in the assumed
uncertainties in the A, B, and C flux measurements, would be necessary
to bring our best-fit $\chi^2/\nu$ values to unity.  In addition, the
lens model is very simple and depends upon the assumptions that the
galaxies share the same redshift, and that there are no significant
additional perturbations along the line of sight.

Changing the variation of mass with radius changes the time delay
between the three bright images and D.  For this reason we believe
that the B-D time delay is more uncertain than Table~4 suggests,
though more simulations are necessary to verify this assertion.

(The ambitious reader may notice that our model produces two spurious
images near the centres of G2 and G3.  These images are only artifacts
of our treatment of G2 and G3 as point masses, an approximation valid
only far from the centres of G2 and G3.  In fact, the actual mass
distributions of G2 and G3 are spread out so that they do not exceed
the critical density necessary to produce additional images of the
quasar.)

\subsection{Lens properties}

In the refined model, the equivalent velocity dispersion of G1 is
$210$~\kms\ and the mass within the critical radius is
$9.3\times10^{10}h^{-1}~M_{\odot}$, consistent with a single normal
galaxy.  Galaxies G2 and G3 have masses
$4.1\times10^{12}h^{-1}~M_{\odot}$ and
$4.7\times10^{11}h^{-1}~M_{\odot}$ respectively.  These numbers were
calculated assuming $\Omega_0=1$ and $z_{\rm d}=0.64$.  If the
redshift is significantly smaller than 0.64, the masses of G2 and G3
need not be as large to produce the same effect in the model
(Browne, personal communication).

These results are consistent with the predictions of the initial
model.  This provides an argument against the existence of a large
density of ``dark'' galaxies as gravitational lenses ought to sample
all the mass in the universe fairly.

\subsection{VLBI observations}

The four images of \B\ have a combined flux density of $\sim0.5$~Jy.
It should therefore be mappable by very long baseline interferometry
({\sc vlbi}).  A {\sc vlbi} map that resolves the bright components
will allow for more detailed models because magnification tensors, not
just magnifications, may then be included in the fit.

If the source has a simple core-jet structure with superluminally
moving features, then it should be possible to test the hypothesis
that the core is stationary.  Furthermore, the strong derived
tangential (\ie\ along the A-B-C arc) magnification in our models
almost guarantees that the observed motion in the three bright images
will also be tangential (with B's motion reversed).  The absolute
expansion speeds depend upon the unknown intrinsic value but it could
exceed $\sim100c$ or $\sim9$~mas\,yr$^{-1}$.

\subsection{Time delay}

If additional information can be obtained from radio observations,
then it will remove some of the uncertainty in the shape of the time
delay surface and allow a moderately accurate prediction of the
relative time delay of images A and C with respect to image B, now
anticipated to be several hours. (That image B's variation should
follow that of images A and C and precede that of image D is mandated
by the topology of the arrival time surface.)  It is possible that
delays this short can be measured if the source is an intraday
variable (\eg\ Quirrenbach \etal\ 1991).  In this case, it will be
possible to fix the multiplicative constant $T_0$ in the time delay
and make a measurement of the Hubble constant.  It may be easier to
measure the time delay between images D and B.  Measurement of relative
magnification tensors will also help as they will remove some of the
uncertainty in the shape of the time delay surface.

Even if either of these approaches furnishes a measurement of the
Hubble constant, there will still be a residual uncertainty associated
with the specific choice of world model and the possibility that a
large screen covers the source. For example, changing $\Omega_0$ to
$\sim0$, while retaining $z_d=0.64$, increases $T_0$ by 25 percent.
Introducing a cosmological constant while retaining flatness for
$\Omega_0=0.1$ (conforming with model C of Carroll, Press \& Turner
1992) increases $T_0$ by 12 percent.

More speculatively, Gott, Park \& Lee (1989) used the observations of
Q2016+112 to exclude the possibility of an antipode at large redshift.
This argument is much stronger in the case of \B\ because we can
exhibit a successful model and the source is known to be at a higher
redshift.

\section{Conclusions}

In this paper, we have exhibited a simple gravitational lens model for
the radio-loud quasar \B\ despite the unprecedented, large relative
magnification of $\sim50$. That this is possible, and that the quality
of the fit improved after the three galaxies were discovered, supports
Patnaik \etal's (1992) and Lawrence \etal's (1992) claim that this
source is multiply imaged.  The primary lensing galaxy G1 is located
within the circle of images.  The galaxies G2 and G3 act
perturbatively on the image arrangement, displacing image D closer to
G1 (and demagnifying its flux in the process) and having the opposite
effect on the other three images.  \B\ provides a prime example of
magnification bias.  At both radio and visual frequencies, the source
has uncommonly large flux.  The combined magnification on the basis of
the refined model is $\Sigma_i|\mu^{(i)}|=29$.  The source quasar is
therefore only $20^{\rm m}$ and the {\em a posteriori\/} probability
of its having been multiply imaged is then less striking as there is a
much larger parent population.

As the source is very bright at radio wavelengths it should be
possible to obtain high dynamic range maps of images A, B, and C. We
predict that all three images are elongated along the A-B-C arc, and
that the parity of image B should be opposite to the parities of
images A and C.  We also predict large proper motions for the
components along the A-B-C arc.  Detailed {\sc vlbi} maps that allow
measurement of the relative magnification tensors may also lead to
more detailed and accurate lens models.

Our final quantitative prediction is the time delay between image
variations.  If \B\ turns out to be a strong intraday variable, then
it may be possible to measure this for the three bright images.
(Success in this endeavor would also provide a very clean proof that
intraday variability is intrinsic to compact radio sources and is not
imprinted by refractive interstellar scintillation.) Relatively large
flux changes in the source will be necessary to measure the predicted
$\sim1$ month delay in the variation of the D image.  Time alone will
tell if the four images of B1422+231 are formed by a simple enough
potential and if the source has sufficient intrinsic variability to
furnish a useful measurement of the Hubble constant.

\section*{Acknowledgements}

We thank Ian Browne (our referee), Erica Ellingson, Chris Kochanek,
Walter Landry, James Larkin, Charles Lawrence, Gerry Neugebauer, Alok
Patnaik, Tony Readhead, Peter Schneider, Tom Soifer, and Howard Yee
for valuable discussions and encouragement.  Support under the NSF
Graduate Fellowship program and NSF grants AST 89-17765 and AST
92-23370 is gratefully acknowledged.

\section*{Note added in proof}

Recent radio observations by Browne (personal communication) confirm
the A/B flux ratio which our model is unable to fit.  M. Remy et al.\
(1993) have observed the \B\ system in several optical bands.  Their
observations are consistent with the morphology found in the radio and
IR, and they also find galaxies G2 and G3.  Because our position for
G1 was measured from a preliminary sketch provided by Yee \&
Ellingson, the position we used for G1 differs somewhat from the
position they have published.  The difference is not significant.
Kormann, Schneider \& Bartelmann (preprint) have shown that the masses
of G2 and G3 can be reduced if G1 is given an ellipticity.

\section*{References}
\begin{list}{}{
\setlength{\itemindent}{-1.0\parindent}
\setlength{\listparindent}{\parindent}
\setlength{\itemsep}{0.0\itemsep}
}
\raggedright

\item
Blandford, R. D., 1990, QJRAS, 31, 305

\item
Blandford, R. D., Narayan, R., 1992, ARA\&A 30, 311

\item
Carroll, S. M., Press, W. H., Turner, E. L., 1992, ARA\&A, 30, 499

\item
Gott, J. R., Park, M.-G., Lee, H. M., 1989, ApJ, 338, 1

\item
Hammer, F. \etal, 1993, poster at 31st Li\`ege International
Astrophysical Colloquium

\item
Hogg, D. W., Blandford, R. D., 1993, BAAS, 25, 794

\item
Kayser, R., 1990, ApJ, 357, 309

\item
Kochanek, C. S., 1991a, ApJ, 373, 354

\item
Kochanek, C. S., 1991b, ApJ, 382, 58

\item
Larkin, J. T. \etal, 1993, in preparation

\item
Lawrence, C. R., Neugebauer, G., Weir, N., Matthews, K., Patnaik, A.
R., 1992, MNRAS, 259, 5{\sc p}

\item
Patnaik, A. R., Browne, I. W. A., Walsh, D., Chaffee, F. H., Foltz, C.
B., 1992, MNRAS, 259, 1{\sc p}

\item
Press, W. H., Teukolsky, S. A., Vetterling, W. T., Flannery, B. P.,
1992, Numerical Recipes in Fortran: The Art of Scientific
Computing, 2ed. Cambridge University Press, Cambridge

\item
Quirrenbach, A. \etal\, 1991, ApJ, 372, L71

\item
Remy, M., Surdej, J., Smette, A., Claeskens, J.-F., 1993, A\&A, 278,
L19

\item
Schneider, P., Ehlers, J., Falco, E. E., 1992, Gravitational Lenses.
Springer-Verlag, New York

\item
Yee, H. C., Ellingson, E. E., 1993, AJ, in press

\end{list}


\newpage
\section*{Tables}
\begin{list}{}{
\setlength{\listparindent}{\parindent}
\setlength{\itemindent}{0.0\parindent}
\setlength{\itemsep}{5.0\itemsep}
}
\item
{\bf Table 1.}  Image positions.
\begin{center}
\begin{tabular}{cccc}
  image & \multicolumn{1}{c}{$\theta_1$} &
\multicolumn{1}{c}{$\theta_2$} &
\multicolumn{1}{c}{$|\mu^{(iB)}|$}\\  \hline
A & $-0.39\pm0.02$ & $0.32\pm0.02$   & $0.98\pm0.02$ \\
B & $0.00$         & $0.00$          & $1.00$ \\
C & $0.33\pm0.02$  & $-0.75\pm0.02$  & $0.52\pm0.02$ \\
D & $-0.94\pm0.03$ & $-0.81\pm0.03$  & $0.020\pm0.005$ \\
\end{tabular}
\end{center}

Relative positions and 5~GHz radio fluxes of the components of \B\
(Patnaik \etal\ 1992).  The coordinates $\theta_1$, $\theta_2$,
advance in the directions of decreasing RA and increasing declination
respectively and are measured in arcseconds.

\newpage
\item
{\bf Table 2.}  Galaxy positions.

\begin{center}
\begin{tabular}{ccc}
  galaxy & \multicolumn{1}{c}{$\theta_1$} &
\multicolumn{1}{c}{$\theta_2$} \\  \hline
G1 & $-0.70$ & $-0.59$ \\
G2 & $-9.0$ & $-5.2$ \\
G3 & $-3.6$ & $-7.3$ \\
\end{tabular}
\end{center}

Positions of the observed lensing galaxies G1 (Yee \& Ellingson 1993),
G2, and G3 (Larkin \etal\ 1993), in the same coordinate system as the
data in Table~1.

\newpage
\item
{\bf Table 3.}  Model parameters.

\begin{center}
\begin{tabular}{cc|cc}
\multicolumn{2}{c|}{initial model} & \multicolumn{2}{c}{refined model} \\
\hline
$b^{(1)}$           & $ 0.62\pm0.02$ & $b^{(G1)}$ & $0.75\pm0.01$ \\
$\theta_{c1}^{(1)}$ & $-0.67\pm0.03$ & $b^{(G2)}$ & $5.0\pm0.7$ \\
$\theta_{c2}^{(1)}$ & $-0.60\pm0.02$ & $b^{(G3)}$ & $1.7\pm0.3$ \\
$b^{(2)}$           & $ 3.02\pm0.02$ & & \\
$\theta_{c1}^{(2)}$ & $-6.6\pm0.4$   & & \\
$\theta_{c2}^{(2)}$ & $-5.0\pm0.3$   & & \\
\end{tabular}
\end{center}

Parameters for the initial and refined models with standard computed
errors.

\newpage
\item
{\bf Table 4.}  Future observables.

\begin{center}
\begin{tabular}{ccc}
observable & \multicolumn{1}{c}{initial model}
           & \multicolumn{1}{c}{refined model} \\ \hline
\\
$[\mu^{(AB)}]$&
$\left[\begin{array}{rr}
 3.2 & 3.5 \\
 -0.81 & -1.1 \end{array}\right] $ &
$\left[\begin{array}{rr}
  2.89 &  3.18 \\
 -0.88 & -1.23 \end{array}\right] $
\\
$[\mu^{(CB)}]$&
$\left[\begin{array}{rr}
 0.27 & 0.32\\
 4.0 & 2.9 \end{array}\right] $ &
$\left[\begin{array}{rr}
 0.34 & 0.37 \\
 3.57 & 2.62 \end{array}\right] $
\\
$[\mu^{(DB)}]$&
$\left[\begin{array}{rr}
 0.65 & 0.53 \\
 0.38 & 0.37 \end{array}\right] $ &
$\left[\begin{array}{rr}
 0.64 & 0.53 \\
 0.33 & 0.32 \end{array}\right] $
\\
\\
$\mu^{(AB)}$ & $-0.7$ & $-0.77$ \\
$\mu^{(CB)}$ & $-0.5$ & $-0.42$ \\
$\mu^{(DB)}$ & $0.04$ & $0.03$ \\
\\
$\Delta t^{(AB)}$ & $-0.2h^{-1}$ & $-0.16h^{-1}\pm0.04$ \\
$\Delta t^{(CB)}$ & $-5h^{-1}$   & $-1.0h^{-1}\pm0.1$   \\
$\Delta t^{(DB)}$ & $+18h^{-1}$  & $+29.9h^{-1}\pm0.8$   \\
\end{tabular}
\end{center}

The computed magnification tensors $[\mu^{(iB)}]$ (and magnifications
$\mu^{(iB)}$) relative to image B, and time delays $\Delta t^{(iB)}$
in days after B, adopting the fiducial value
$T_0=62h^{-1}$~d\,(arcsec)$^{-2}$.

\end{list}

\newpage
\section*{Figures}
\begin{list}{}{
\setlength{\itemindent}{0.0\parindent}
\setlength{\listparindent}{\parindent}
}

\item
{\bf Figure 1.} Time delay surface for the refined model.  The images
are located at A, B, C, D and elongated along the same directions as
the contours.  Contours are separated by about 1.1$h^{-1}$ days.

\end{list}

\end{document}